\RequirePackage{fix-cm}
\documentclass[smallcondensed]{svjour3}     % onecolumn (ditto)
\smartqed  % flush right qed marks, e.g. at end of proof
\usepackage{graphicx}
\usepackage{url}
\usepackage{color}
\newcommand{\jasr}{    {\it Adv. Space Res. }} 
 
\newcommand{\araa}{   {\it Ann. Rev. Astron. Astrophys. }}
\newcommand{\aap}{    {\it Astron. Astrophys. }}
\newcommand{\aaps}{   {\it Astron. Astrophys. Suppl. }}
\newcommand{\aapr}{   {\it Astron. Astrophys. Rev. }}

\newcommand{\apj}{    {\it Astrophys. J. }}
\newcommand{\apjl}{   {\it Astrophys. J. Lett. }}

\newcommand{\grl}{    {\it Geophys. Res. Lett. }}

\newcommand{\jgr}{    {\it J. Geophys. Res. (Space Phys.)}}

\newcommand{\solphys}{{\it Solar Phys. }}
 
\newcommand{\ssr}{    {\it Space Sci. Rev.}}

\journalname{Space Science Reviews}
\begin{document}

\title{Acceleration and propagation of Solar Energetic Particles%\thanks{Grants or other notes
%about the article that should go on the front page should be
%placed here. General acknowledgments should be placed at the end of the article.}
}
%\subtitle{Do you have a subtitle?\\ If so, write it here}

\titlerunning{Solar Energetic Particles}        % if too long for running head

\author{Karl-Ludwig~Klein         \and
        Silvia~Dalla
}

%\authorrunning{Short form of author list} % if too long for running head

\institute{K.-L.~Klein \at
            LESIA - Observatoire de Paris, CNRS (further affiliations: PSL Research University, Univ. P \& M Curie, Univ. Paris-Diderot), Meudon, France\\
             \email{ludwig.klein@obspm.fr}
         \and
              S.~Dalla \at
              Jeremiah Horrocks Institute \\
              University of Central Lancashire, UK \\                   
              \email{sdalla@uclan.ac.uk}           %  \\
%             \emph{Present address:} of F. Author  %  if needed
            } 

\date{Received: date / Accepted: date}
% The correct dates will be entered by the editor

\maketitle

\begin{abstract}

Solar Energetic Particles (SEPs) are an important component of Space Weather, including radiation hazard to humans and electronic equipment, and the ionisation of the Earth's atmosphere. We review the key observations of SEPs, our current understanding of their acceleration and transport, and discuss how this knowledge is incorporated within Space Weather forecasting tools. Mechanisms for acceleration during solar flares and at shocks driven by Coronal Mass Ejections are discussed, as well as the timing relationships between signatures of solar eruptive events and the detection of SEPs in interplanetary space. Evidence on how the parameters of SEP events are related to those of the parent solar activity is reviewed and transport effects influencing SEP propagation to near-Earth locations are examined. Finally, the approaches to forecasting Space Weather SEP effects are discussed. We conclude that both flare and CME shock acceleration contribute to Space Weather relevant SEP populations and need to be considered within forecasting tools.

%Include keywords, PACS and mathematical subject classification numbers as needed.
% \PACS{PACS code1 \and PACS code2 \and more}
% \subclass{MSC code1 \and MSC code2 \and more}
\end{abstract}

%-----------------------------------

%\section{Introduction}

\section{Introduction}
%\subsection{}

Solar energetic particle (SEP) events are distinct enhancements in space of particle fluxes of electrons, protons and heavy ions at energies well above the average thermal energy in the corona, which is a few hundreds of eV. The energy of solar energetic protons can reach GeV in some events. The events last from a few hours to several days. Space weather effects of SEPs are due to the interaction of the particles with electronics, the Earth's atmosphere and living beings. Strong SEP events are a potential space weather hazard because they may affect space borne electronics and generate substantial particle radiation. Energetic ions are themselves primary radiation and they generate secondaries through nuclear reactions with the Earth's atmosphere and any exposed materials. This leads to enhanced ionisation and sometimes to modifications of the local chemistry of the high polar atmosphere of the Earth. SEP events are a major obstacle to human spaceflight outside the Earth's magnetosphere.

In this paper we discuss the acceleration and propagation of  SEPs from a space weather perspective. Scientific understanding of these topics underpins many forecasting tools that aim to predict the occurrence of SEP events and their impact.

Two key questions regarding SEPs within a Space Weather framework are: Which solar events are responsible for the production of SEPs most dangerous for Space Weather? What type of events can give rise to extreme SEP enhancements? In this chapter we review the observational evidence and modelling efforts related to SEPs, with the aim of addressing the above questions.

Since SEP events usually have well-defined onset time and intensity enhancements up to several orders of magnitude, they can be related to individual events of solar activity. Historically flares and filament eruptions were the only known major transient features in the solar atmosphere. SEP events were considered as a consequence of flares until the 1970s. Since the discovery of CMEs in the 1970s a bimodal framework for the interpretation of SEP events has emerged, based on their classification into impulsive and gradual \cite{Rea1999}. Impulsive SEP events are short ($\leq$1~day), low intensity and numerous (estimated as about 1000/year in periods of high activity). Gradual events are long (several days at energies of a few MeV/nuc), rather rare (a few tens per year), and orders of magnitude more intense in protons than impulsive SEP events. Within this scenario, the latter type of events, characterised by the largest proton fluences and therefore of most relevance to Space Weather, are ascribed to acceleration by CME-driven shocks as they propagate through the heliosphere. There is some debate as to the role played by ``flare acceleration"  in these events. The question for space weather forecasting is whether it suffices to consider the parameters of CMEs in order to provide an accurate SEP forecast, or whether flare information must also be taken into account. One can immediately see that the science related to SEP acceleration should inform Space Weather, for example through statistical studies of links between solar events and SEP occurrence and parameters. 

The propagation of energetic particles through interplanetary space is also key to determining whether or not SEPs will be observed, for example, near Earth and potentially produce Space Weather effects. The classic description of SEP transport, embedded within the two-class paradigm, is spatially one-dimensional, in the sense that particles are assumed to remain tied to the magnetic field line on which they were initially injected. The detection of particle flux at one location therefore requires direct magnetic connection to the acceleration region. At the same time, measurements from a number of spacecraft have shown that in many events particles have easy access to wide regions in longitude and latitude. Is this observation the result of wide CME driven accelerating fronts in the heliosphere, as postulated within the two-class paradigm, or could other physical processes be at play? 
Again scientific understanding is required to assess and forecast the Space Weather effects of SEPs.

We begin in Section \ref{sec.keyobs} by reviewing key SEP observations, on which our current understanding is based and which need to be explained by any overarching SEP theory. In Section \ref{sec_accel} we briefly touch upon mechanisms for particle acceleration that are thought to operate at the Sun, and in Section \ref{sec_timing} we discuss the links and timings between remote sensing signatures of acceleration and SEP detection in interplanetary space. Section \ref{sec_stat} summarises studies of statistical relationships between solar activity and SEP occurrence and parameters. In Section  \ref{sec.transp} the particle transport through the corona and interplanetary space is discussed. Approaches to forecasting SEPs within a Space Weather perspective are reviewed in Section  \ref{sec.forecast} and finally Section \ref{sec.disc} presents our conclusions and discusses future perspectives. 

%%%%%%%%%%%%%%%%%%%%%%%%%%%%%%%%%%%%%%% Key observations %%%%%%%%%%%%%%%%%%%%%%%%%%%%%%%%%%%%%%%%%%%%%%%

\section{Key observations} \label{sec.keyobs}

A large amount of observations have been gathered on SEPs over the past decades. It is beyond the scope of this article to review all available experimental information, but in this Section we will focus on the main features of SEP observations. The reader is referred to reviews by Reames \cite{Rea1999}, Klecker {\it et al} \cite{Klc:al-06} and Desai and Giacalone \cite{Des2016} for additional information.

\begin{figure}[htbp] 
   \centering
   \includegraphics[width=0.45\textwidth]{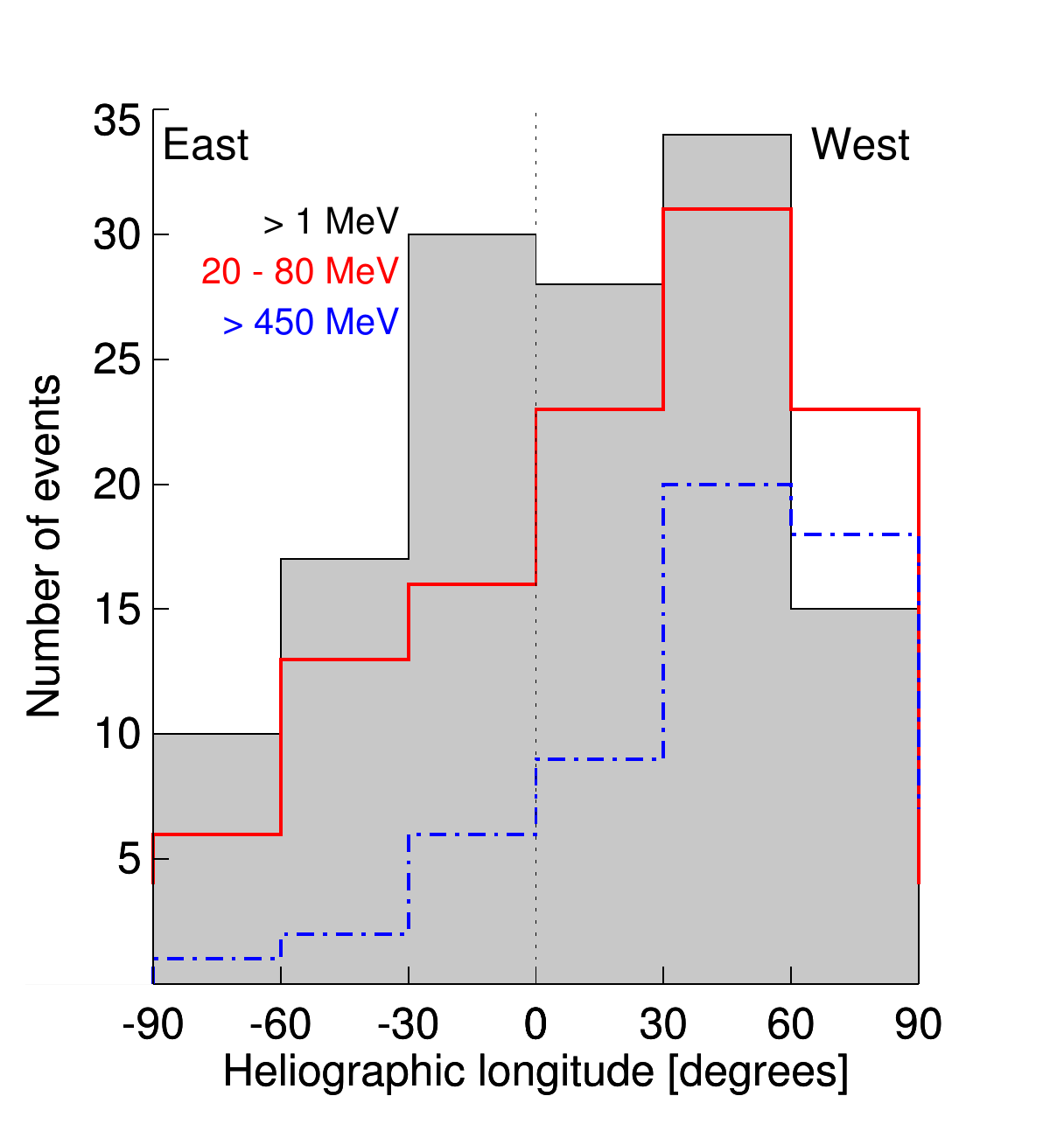} 
   \includegraphics[width=0.45\textwidth]{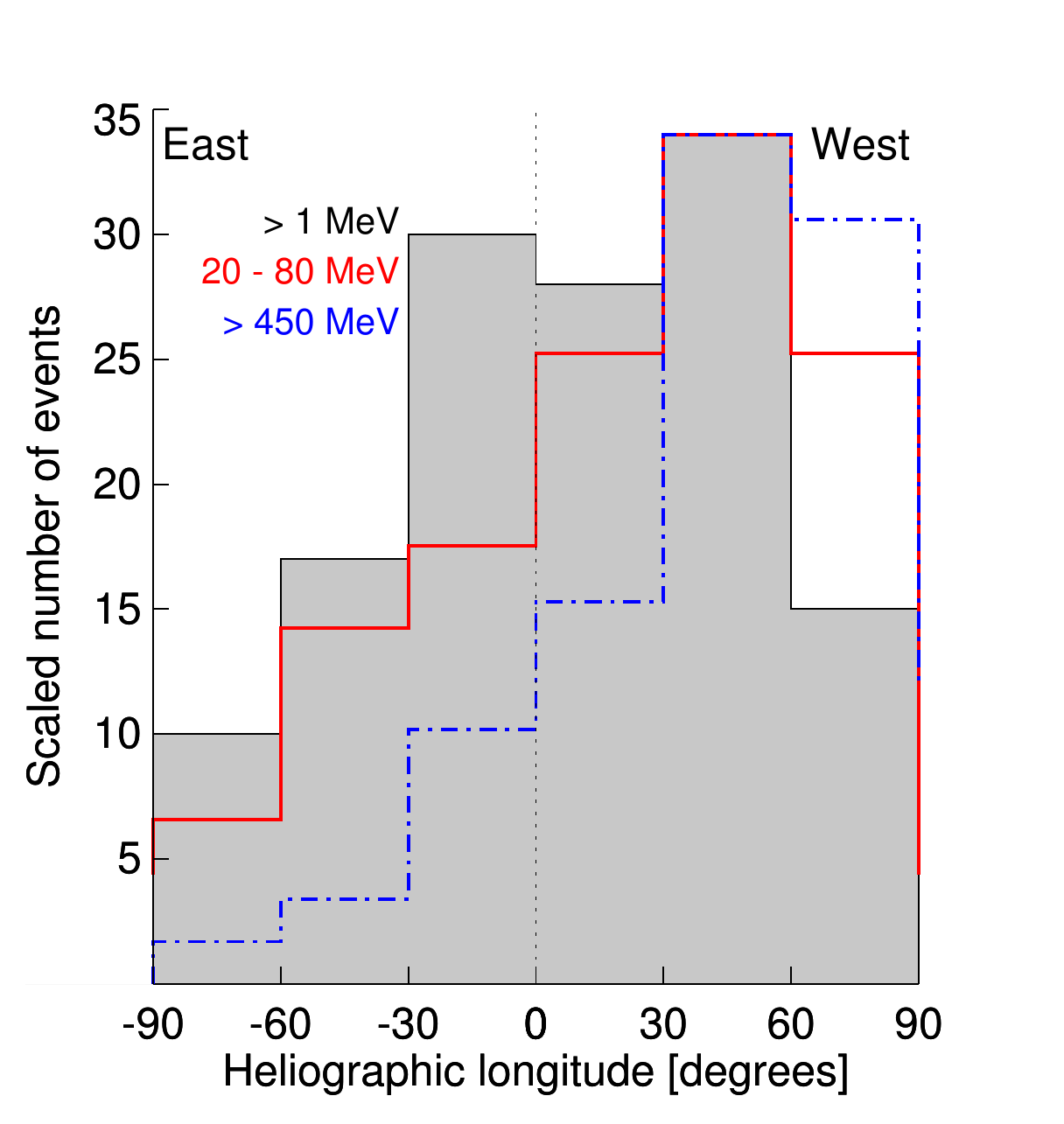}

\vspace{-2ex}
\noindent { (a) \hspace{0.48\textwidth} (b) \hspace{0.4\textwidth} \mbox{} }
   \caption{Heliolongitude distribution of the flares associated with SEP events in different energy ranges: gradual events at energies $>$1~MeV (black line and grey background shading; numbers rebinned into 30$^\circ$-wide intervals, after Fig.~2.3 of \cite{Rea1999}); 20-80~MeV (red solid line) from  \cite{vHo1975}; $>$450~MeV (blue dashed-dotted line) from the list of ground level enhancements (GLEs) at www.nmdb.eu. Panel (a) gives the actual event numbers, (b) the numbers scaled to the maximum of the distribution of $>$1~MeV events.
   }
   \label{Fig_ldist}
\end{figure}

\subsection{Location of parent active region}

SEP enhancements associated with solar flares and Coronal Mass Ejections (CMEs) can last from just a few hours to several days. Whether or not an SEP event will be detected, for example near Earth, depends on the location at the Sun of the parent event: eruptions at Western locations on the solar disk have a much higher likelihood of resulting in an SEP event at Earth, due to the curvature of the Parker spiral interplanetary magnetic field. However, in many events particles can be detected at locations widely separated, either in longitude ({\it e.g.},~\cite{Ric:al-14}) or in latitude ({\it e.g.},~\cite{Dal2003b}), from the parent region. While this was thought initially to apply only to gradual events, it has been shown that impulsive events can also be detected over wide longitudinal ranges \cite{Wie2013}. In some cases particles can spread over 360$^{\circ}$ in longitude \cite{Gom2015}.

Observed longitude distributions of the flares associated with major SEP events are displayed in Figure~\ref{Fig_ldist}, for three separate SEP energy ranges. The black histogram and grey shading show gradual SEP proton events at $\sim$1 MeV \cite{Rea1999} (where numbers have been rebinned into 30$^{\circ}$-wide intervals for the sake of comparison). The red line shows the histogram for the proton energy range 20-80 MeV \cite{vHo1975} and the blue dashed-dotted line that for protons above 450 MeV (the nominal atmospheric cutoff energy of neutron monitors at sea level). All three distributions peak in the nominally well-connected longitude range W30$^{\circ}$--W60$^{\circ}$. But all have a significant width, showing that particles can reach Earth-connected interplanetary field lines even when the associated flare is several tens of degrees away from the footpoint of the nominal Parker spiral on the solar wind source surface. The width of the distribution decreases with increasing particle energy: gradual SEP events detected at MeV energies have a very weak longitude dependence of the associated flare, while GLEs have a more pronounced concentration around the well-connected field lines. An intermediate behaviour is displayed by SEP events in the 20-80~MeV range.

\subsection{Particle intensities and anisotropies}

Profiles of particle intensities versus time vary depending on the location of the parent event, with SEPs from Western events displaying a fast rise and generally shorter duration, and those from Eastern events having a much more gradual rise phase \cite{Can1988}. Peak intensities show a strong dependence on the longitudinal separation from the source active region and tend to be largest at the spacecraft best connected to the region ({\it e.g.},~\cite{Ric:al-14}). In some events intensities show a large peak at energies up to $\sim$100~MeV at the time when a CME-driven interplanetary shock passes the Earth. The phenomenon is termed Energetic Storm Particle (ESP) event. ESP events are not addressed here in detail. Recent reviews of observations and theory are given in  \cite{Lee:al-12} and \cite{Des2016}. 

Anisotropies, describing the degree to which particles are beamed when arriving at the detecting spacecraft, are an important property of SEP events that can be obtained from sectored particle instruments. Anisotropies tend to be larger at the beginning of an event, and usually they quickly fall to zero, indicating isotropy of the particle distribution. In a recent study of electron anisotropies detected by STEREO, it was shown that in many events anisotropies are large at the spacecraft best connected to the flare region while they are small at less well connected spacecraft, although other types of multi-spacecraft anisotropy signatures are also possible \cite{Dre2014}.

The energy range over which SEPs are observed at 1 AU varies considerably between different events and is of course dependent on instrumentation and background fluxes. In the largest events, the presence of protons of energies up to tens of GeVs can be detected through neutron monitors at the Earth's surface, in so-called Ground Level Enhancements (GLEs). Measurements of solar $\gamma$-rays by the FERMI/LAT instrument have shown that the acceleration of protons of energies above 300 MeV is a common occurrence in the solar corona (see Sect.~\ref{sec_timing_FERMI}) and can even be observed in M-class flares 

\subsection{Elemental abundances and charge states}

Measured properties of heavy ion SEP populations provide important information on their acceleration and propagation. Both depend on the mass-to-charge ratio $m/q$. Heavy ion SEPs detected at 1 AU are typically partially ionised.

Heavy ion properties provide some key criteria for separating events into gradual and impulsive within the two-class scenario \cite{Rea1999}. In the initial version of this scenario impulsive events are characterised by values of the event-averaged Fe/O around 1 (about 10 times larger than typical coronal values), while gradual events are less rich in iron, with values around 0.1. Impulsive events show large enhancements in the event-averaged $^3$He/$^4$He ratio (about 1000 times larger than typical coronal values), while the ratio is typically orders of magnitudes lower in gradual events. Charge states of Fe in impulsive events are typically higher ($Q_{Fe}$$\simeq$20) than in gradual events ($Q_{Fe}$$\simeq$12)  \cite{Rea1999}.

\begin{figure}[htbp] 
   \centering
   \includegraphics[height=5.5cm]{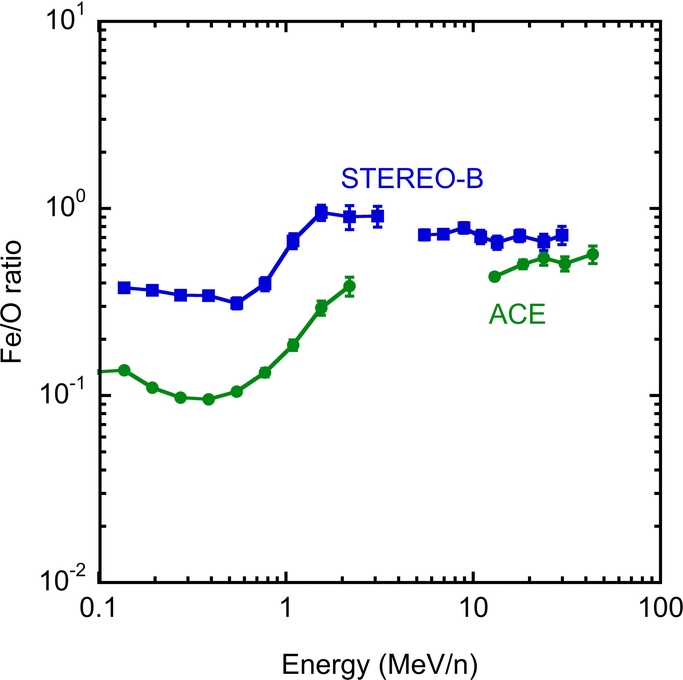} 
   \includegraphics[height=5.8cm]{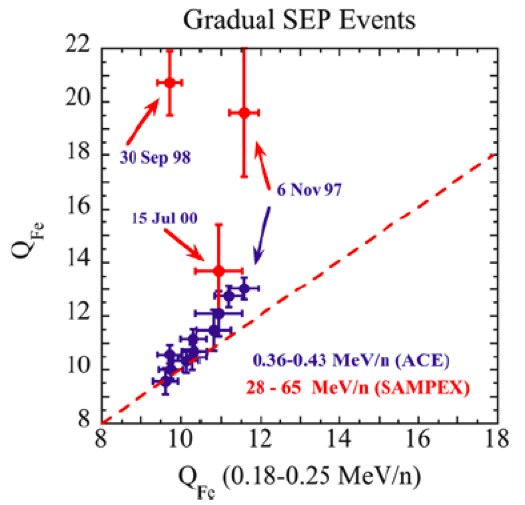} 

\vspace{-2ex}
\noindent { (a) \hspace{0.48\textwidth} (b) \hspace{0.4\textwidth} \mbox{} }
      \caption{(a) Energy dependence of Fe/O ratio for the event of 2013 April 11 \cite{Coh2014}. \copyright AAS. Reproduced with permission. (b) Compilation of energy dependence of Fe charge states for a number of SEP events \cite{Klc:al-07}. \copyright Springer. Reproduced with permission. 
}
   \label{energy_dep}
\end{figure}

The above criteria were established in the1990s using SEP data for heavy ions in the $\sim$1 MeV/nuc energy range. It was realised later that charge states are in fact strongly dependent on the energy of the ions. In the common impulsive SEP events, charge state at energies below 1~MeV/nuc were found to increase with increasing energy \cite{Klc:al-07,DiF:al-08}. In several large (\lq gradual\rq)  events the charge states were shown to increase with energy above 1~MeV/nuc such as to approach values at several tens of MeV/nuc that were considered as typical of \lq impulsive\rq  $\;$ SEP events in the 1990s \cite{Coh1999,Maz1999,Moe1999,Mew2006,Klc:al-07,Rll:al-12}. The left-hand panel in Figure~\ref{energy_dep} shows the Fe/O ratio during an event as a function of particle energy, observed by two spacecraft at different heliolongitudes.  At both spacecraft the Fe/O ratio increases with increasing energy above $\sim$0.7~MeV/nuc. In addition, the ratio is different at different vantage points. The right-hand panel shows event-integrated Fe charge states in two energy ranges (blue and red crosses, respectively) as a function of charge states in the energy range 0.18-0.25 MeV/nuc for several gradual SEP events. Each cross designates a different event. Events with energy-independent charge states are on the inclined dashed line. The blue crosses are mostly close to  this line, showing that charge states in the ranges (0.36-0.43) MeV/nuc are the same as those in the (0.18-0.25)~MeV/nuc range. However, at energies above 20~MeV the average charge of the Fe ions is significantly higher, with charge states typical of the classical value for impulsive events on 1998 Sep 30 and 1997 Nov 06. Elemental abundances also display a time variation in individual events. For example the Fe/O ratio often decays over time from values considered typical of impulsive events, to much lower values \cite{Tyl2013,Mas2006}. Other ionic ratios also display a time dependence and this is dependent on their $m/q$ values \cite{Zel2017}.

\begin{figure}[htbp] %  figure placement: here, top, bottom, or page
   \centering{
   \includegraphics[width=0.48\textwidth]{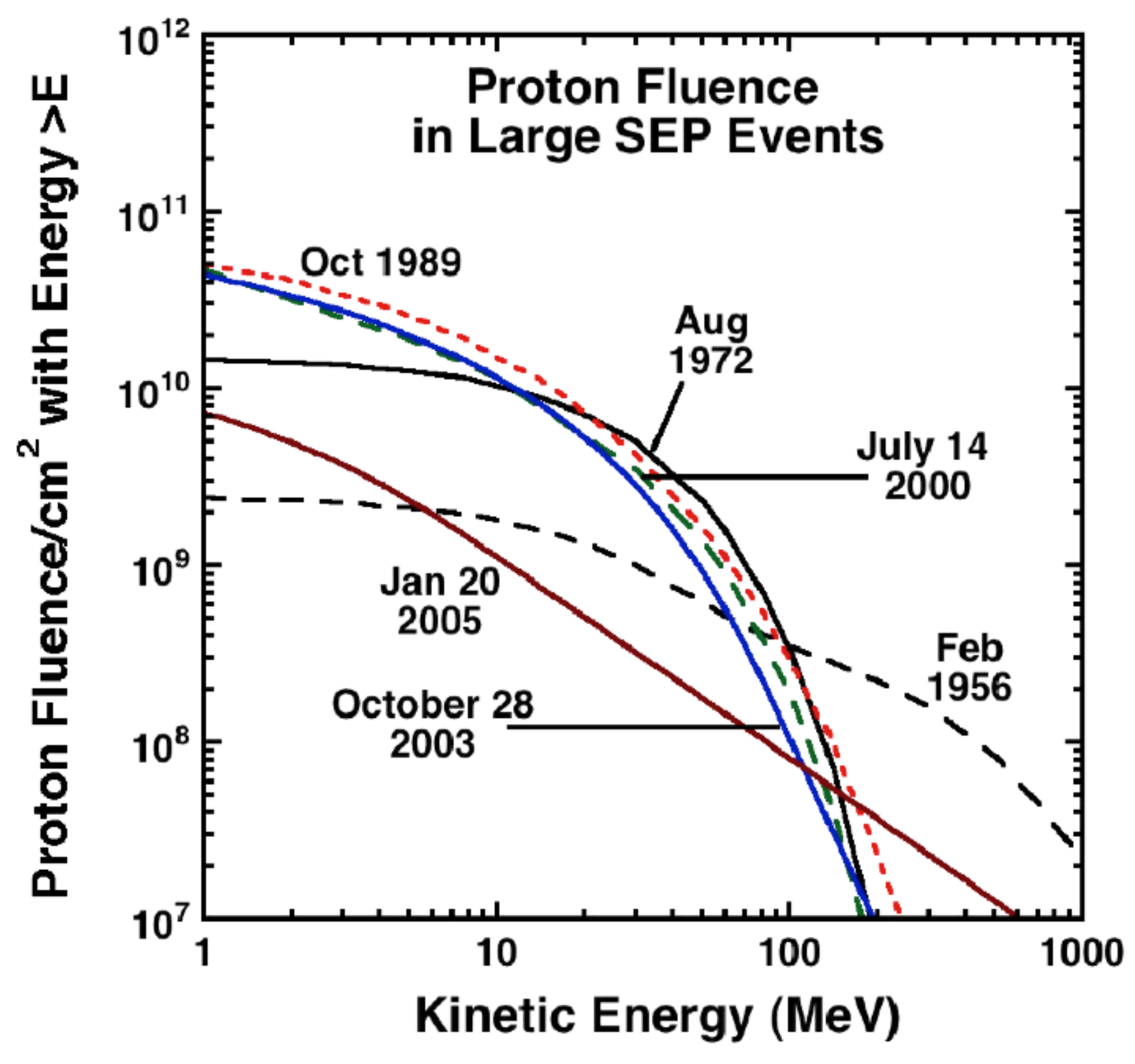} 
    \includegraphics[width=0.48\textwidth]{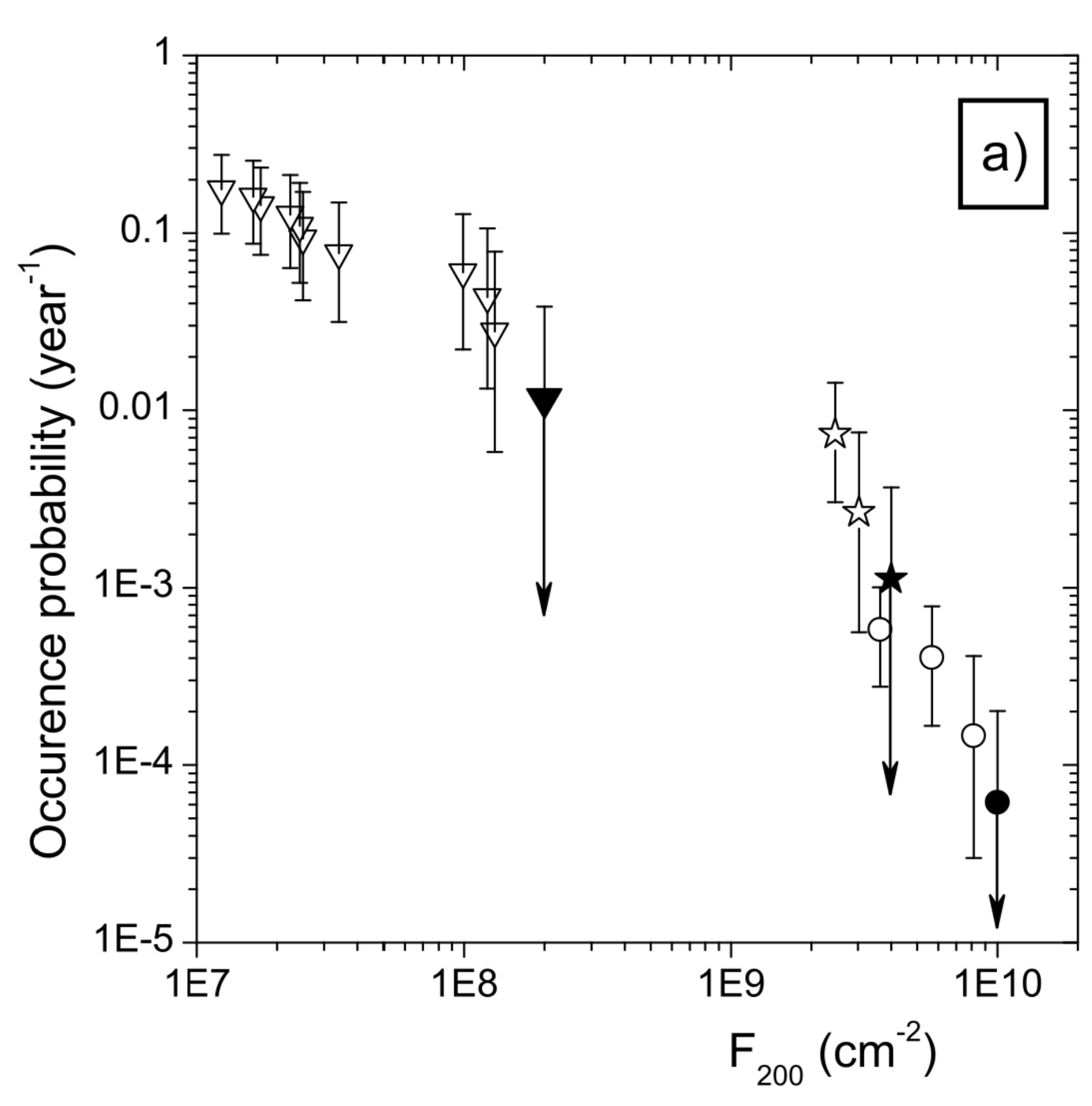} 
}%
\vspace{-2ex}
\noindent { (a) \hspace{0.48\textwidth} (b) \hspace{0.4\textwidth} \mbox{}  \\}
   \caption{(a) Fluence spectra of strong SEP events observed from spacecraft (reproduced from \cite{Mew:al-07a}, with the permission of AIP Publishing). (b) Cumulative occurrence probability of yearly fluences at energies above 200 MeV inferred from neutron monitor observations of relativistic SEP events and from radionuclides in ice cores \cite{Kov:al-14}. \copyright Springer. Reproduced with permission. }
   \label{Fig_SEPspec}
\end{figure}

\subsection{Extreme SEP events}

Studies of space weather effects from SEPs can look back to about 50 years of space observations at energies up to a few hundreds of MeV, and to 70 years of ground-based measurements of relativistic SEP events, up to some GeV or even a few tens of GeV. Fluence spectra of strong events are shown in Fig.~\ref{Fig_SEPspec}.a. The spectra can be divided in two categories: (1) events with high fluence at several tens of MeV, with steep spectra towards higher energies (typical examples: August 1972, October 1989, July 2000), and (2) relativistic particle events, called ground-level enhancements (GLEs) that have spectra extending into the GeV range, but are much weaker at energies of tens of MeV than the first category. The distinction is not bimodal, since the events of Aug 1972, Oct 1989 and Jul 2000 also were GLEs. But high particle intensities at tens of MeV and several GeV do not appear to be correlated, and extreme event scenarios must be elaborated separately for the two categories. This is important since the two relevant energy ranges define different potential space weather hazards: impact on satellites in exposed orbits and on ionisation of the high polar atmosphere at tens of MeV, and tropospheric effects through the formation of atmospheric cascades at several GeV. The reason for the difference is that the extreme fluences at tens of MeV are not dominated by acceleration at the Sun, but near Earth, when it encounters a CME-driven interplanetary shock. The high particle fluxes come from the proximity of the accelerator, rather than its intrinsic strength. GeV particles are accelerated closer to the Sun. 

The probability of occurrence of SEP events at energies of tens of MeV has been studied, {\it e.g.}, by \cite{Xap:al-99}. From an analysis of SEP events in cycles 20 to 22, considering only events occurring in a time interval from 2.5 years before to 4.5 years after the sunspot maximum, these authors inferred a power-law with a rollover at high fluences and a smooth approach of zero at some maximum fluence. The extrapolated maximum fluence at energies above 30~MeV is found to be near $1.3 \cdot 10^{10}$~cm$^{-2}$, about 1.6 times the highest fluence observed. There are probably some differences in the calibration with Fig.~\ref{Fig_SEPspec}.a, which shows slightly lower fluences than Fig.~1 of \cite{Xap:al-99}. But this does not affect the above ratio. Worst case scenarios for prescribed mission durations are also presented by these authors. More recent analyses of extreme events relevant to space missions are, {\it e.g.}, \cite{Jig:al-14}.  

The data base for relativistic solar particle events goes beyond the space age. Direct measurements date back to the 1940s, showing events that are much stronger than during the space age \cite{McC-07}. If transient enhancements in the traces of cosmogenic radionuclides in ice cores and tree rings, such as $^{10}$Be and $^{14}$C, can be related to past solar events, the historic view covers the last 10,000 years. This has been exploited by \cite{Kov:al-14} to establish the cumulative occurrence probability of yearly ion fluence at energies above 200~MeV (Fig.~\ref{Fig_SEPspec}.b). The result is that once every 100 years a yearly fluence can be expected of about 5 times or more the strongest yearly fluence observed by a network of neutron monitors so far, that is the fluence in 1956, with the strongest GLE ever seen -- GLE1956Feb23 (GLE 5). This estimate does not depend crucially on the interpretation of signatures in ice cores or tree rings. Those suggest that the distribution function derived from observed GLEs rolls over at a fluence near $2 \cdot 10^9$~cm$^{-2}$, which is about 25 times the 1956 fluence. There is still some debate, however, if the high-fluence branch in the plot is really related to solar events. Furthermore, the fluences are yearly values, not individual events. No signature of the Carrington event in 1859 is identified in these terrestrial archives. 

%%%%%%%%%%%%%%%%%%%%%%%%%%%%%%%%%%%%%%%  Acceleration processes %%%%%%%%%%%%%%%%%%%%%%%%%%%%%%%%%%%%%%% 

\begin{figure}
  \includegraphics[height=5.5cm,clip=]{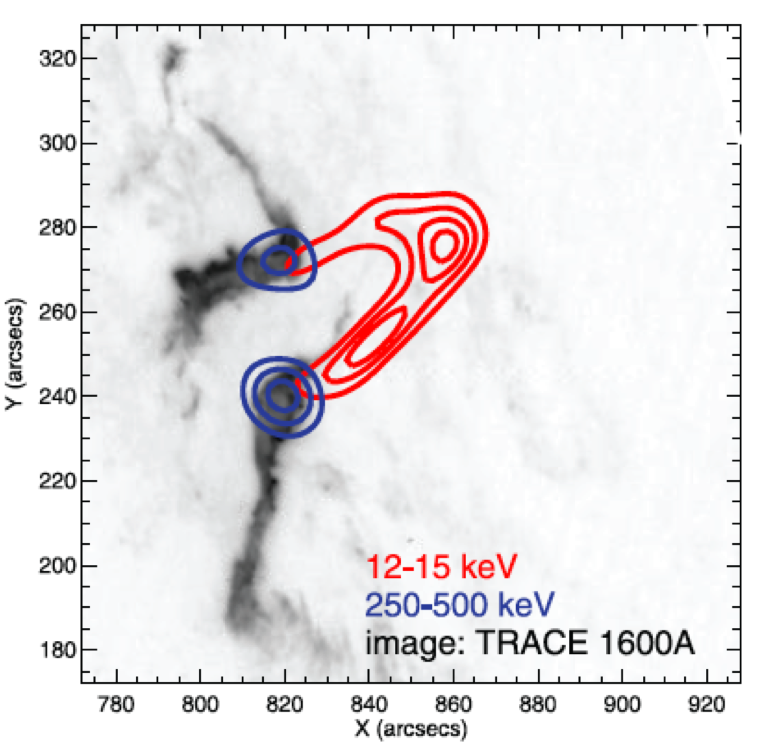}
  \includegraphics[height=5.5cm,clip=]{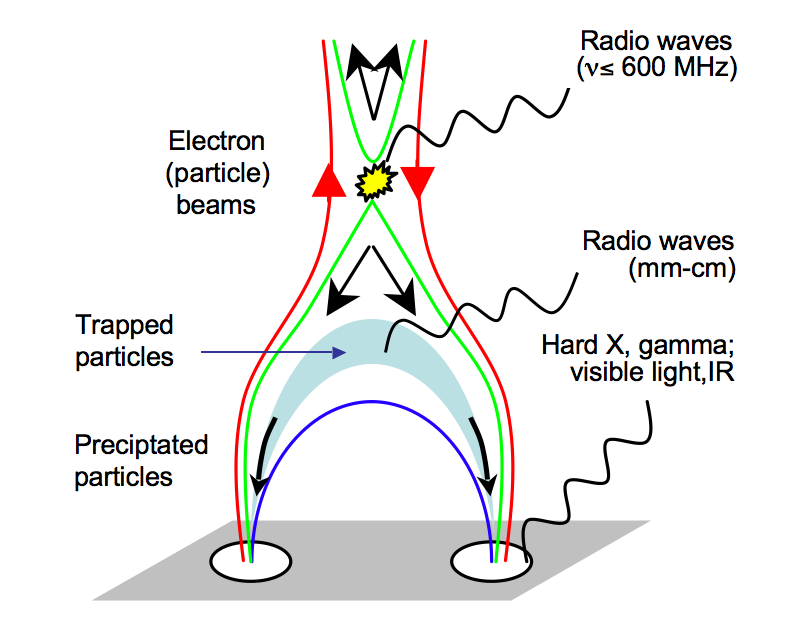}

\vspace{-2ex}
\noindent { (a) \hspace{0.48\textwidth} (b) \hspace{0.4\textwidth} \mbox{} }
\caption{
(a) Contours of hard X-ray emission in two spectral bands (RHESSI) superposed on a TRACE image of chromospheric flare ribbons in UV \cite{Kru:al-08}. \copyright AAS. Reproduced with permission. 
(b) Cartoon scenario of particle populations and related electromagnetic emissions during a flare.
}
\label{Fig_scen}
\end{figure}

\section{Acceleration processes}
\label{sec_accel}

The association of SEP events with flares and CMEs leads to the distinction of two groups of acceleration mechanisms. One is related to magnetic reconnection and turbulence in active regions, exemplified by solar flares and particularly the hard X-ray, gamma-ray and microwave signatures during the impulsive flare phase. The other is particle acceleration at large-scale shock waves, exemplified by interplanetary shocks. Energetic storm particle events accelerated when such a shock passes near Earth are a prime example.

The neat distinction between the two types of processes may be difficult in practice. Magnetic reconnection, for instance, is not restricted to the impulsive flare phase, but may also occur in the downstream region of CMEs, where current sheets are formed, and at the interface between the outward propagating CME and the ambient corona. Shock acceleration may also occur in reconnection regions, at the interface between the reconnection jet and its coronal environment (see \cite{War:al-09a}, and references therein). 

\subsection{Particle acceleration in solar flares}

\subsubsection{Observational evidence on the location of the acceleration region}

Hard X-ray imaging with Yohkoh and RHESSI frequently shows configurations suggesting magnetic loops with a particle acceleration region near or above the loop top. The RHESSI image of the 2005 Jan 20 flare (Figure~\ref{Fig_scen}a) is a prime example: the red contours outline a thermal X-ray source, which traces the upper part of a coronal loop. The blue contours are the sources of hard X-ray emission from the chromospheric footpoints. They project onto flare ribbons seen in UV (the two elongated grey bands), which outline the part of the chromosphere heated by energy deposition during the flare. This source morphology is generally interpreted as a signature of energy release near or above the loop top, which heats the plasma in the coronal loop and accelerates electrons that escape from the primary acceleration site as beams. Because of their high energy they interact very little with the dilute coronal plasma and precipitate into the dense chromospheric footpoints, where they lose their energy instantaneously through collisions with the ambient medium, while simultaneously transforming a small amount of energy into hard X-ray emission. The RHESSI image is a snapshot: during the impulsive phase of the event the X-ray sources occur in an irregular temporal succession at neighbouring places, but the hard X-ray footpoints are always located on the UV ribbons. A standard cartoon scenario is shown in Figure~\ref{Fig_scen}b. The acceleration is ascribed to energy release above the loop top, probably related to magnetic reconnection. The upward field lines may be part of a plasmoid that is ejected upward, or they may be open to the high corona. The energy release may equally well be related to magnetic reconnection with another closed magnetic structure, as in the classical scenario in \cite{Hey:al-77}. The observation \cite{Sui:al-04} of pairs of hard X-ray sources at flare loop tops with temperature gradients pointing to a region between them corroborates the idea that the basic flare energy release occurs near or above loop tops in active regions. 

From the energy-dependent timing of hard X-ray intensity peaks and from the observation that decimetric radio emission shows bidirectional electron beams, with downward-directed beams at high frequency (high ambient electron density) and upward directed beams at low frequency, Aschwanden and coworkers concluded that the typical acceleration region is placed at an altitude of about 1.5 times the half-length of the magnetic loop above the photosphere, with an ambient density of about $(1-10) \cdot 10^9$~cm$^{-3}$ (see Sects. 3.3 and 3.6 of \cite{Asc-02}, and references therein). Recent analyses show that similar bidirectional streams can also show up at higher frequencies, and hence higher ambient electron densities:  \cite{TBL:al-16} derive electron densities in the acceleration region in the range $(1-10) \cdot 10^{10}$~cm$^{-3}$.

An alternative estimate of the thermal electron density in the acceleration region can be obtained from the Fe charge states measured in SEPs, $Q_{Fe}$. High values of $Q_{Fe} \simeq 20$ (to be compared with $Q_{Fe} \simeq 12$ in the quiet corona) are observed in impulsive SEP events and at high energies in some gradual SEP events: if these charge states are to be explained by collisional stripping in the acceleration region, the product of the election density $n_e$ and the residence time in the acceleration region $\tau$ must be $(1-10) \cdot 10^{10}$~cm$^{-3}$~s \cite{Koc:al-00,Krt:al-07}. This is consistent with the high electron densities inferred by \cite{TBL:al-16}, given that the time scales of hard X-ray emission suggest acceleration times $\tau < 1$~s for the electrons. The acceleration time of interacting protons at energies of tens of MeV is unlikely to be significantly higher, since similar time scales of hard X-ray emission and nuclear gamma-ray emission have been found in some flares where the gamma rays were observed with high time resolution \cite{Kan:al-86,Tro:al-93}. 

\subsubsection{Current sheets, magnetic reconnection and particle acceleration}

Magnetic reconnection is widely believed to be the fundamental process explaining both the restructuring of coronal magnetic field configurations and the conversion of the magnetically stored flare energy. Particle acceleration can be due to the direct electric field in the diffusion region, turbulence created during the process, or the reconnection shock that may form as the outflow jets from the diffusion region impinge onto underlying magnetic field structures. These processes are not expected to be neatly separated in a realistic scenario. Current sheets related to solar flares have spatial scales that are largely below the spatial resolution of any imager. They are expected to be highly fragmented and highly dynamic, with the formation and coalescence of magnetic islands, and the retraction of reconnected magnetic field lines \cite{Kli:al-00}. The interaction of particles wandering around within a sea of magnetic islands can lead to more efficient particle acceleration than a single reconnection region as considered in many earlier studies ({\it e.g.}, \cite{Her:al-02}). The fragmentation of current sheets is also physically more realistic than reconnection in a single large-scale structure.  A detailed discussion is given in \cite{Crg:al-12}. Multiple current sheets can also be formed throughout a magnetic loop that is twisted by motions of the photospheric plasma at one of its footpoints \cite{Grd:al-13}. Such a scenario with distributed sites throughout the volume of the loop may resolve problems with understanding the stable transport of intense electron beams over macroscopic distances from the coronal acceleration sites to the chromospheric footpoints where the hard X-rays are emitted. Recent reviews of particle acceleration in solar flares can be found in \cite{Hol:al-11}, \cite{Fle:al-11} and \cite{Vil:al-11}. 

Scenarios of CME eruption include current sheets either as the fundamental driver of the eruption or as a consequence. In coronographic observations bright ray-shaped features that form in the aftermath of the ejected structure are often interpreted as current sheets. The visible  features are dense, and might better be called plasma sheets. But judging from their geometry they must contain current sheets. UV coronographic spectroscopy shows they start hot, at temperatures of several tens of MK, and cool down over time scales of several hours. This is accompanied by the appearance of new loop structures over an extended range of temperatures and wavelengths, and by signatures of sunward-retracting structures, which is reminiscent of the depolarization observed in magnetic substorms. Reviews of recent work are given in \cite{Ray:al-12} and \cite{LnJ:al-15}. These post-eruptive current sheets are likely regions of long-lasting electron acceleration seen in metric radio bursts \cite{Aur:al-09,Ben:al-11,Dml:al-12,Dml:al-12a}, and were suggested to also be a site of  long-lasting acceleration of high-energy ions that escape to the interplanetary space \cite{Car-64,Aki:al-96,Kle:al-14}.

\subsubsection{Stochastic acceleration}

Stochastic acceleration generalises Fermi's idea that charged particles gain energy by encounters with randomly moving magnetic obstacles, because energy-increasing head-on collisions occur more often than energy-reducing trailing collisions.  Instead of macroscopic magnetic obstacles, any fluctuating  electric field in different types of plasma waves and in turbulence can provide the acceleration. A review of stochastic acceleration especially in solar flares is given in \cite{Pet-12}. 

Turbulence is expected to arise in many situations where plasma flows occur. In the solar atmosphere such plasma flows occur in magnetic reconnection, but also at the interface between jets or CMEs and the ambient corona. In a flare scenario as shown by the cartoon in Figure~\ref{Fig_scen}b magnetic energy released during magnetic reconnection or other processes is expected to create high levels of turbulence on MHD scales, which will cascade to smaller scales. Charged particles may be trapped in this turbulence around the loop top, and be accelerated. This could explain why loop top sources are commonly seen in hard X-ray images \cite{Kru:al-08b}. They are especially observed whenever the bright footpoints, which dominate the hard X-ray morphology of flares on the disk, are occulted by the solar disk. 

The cyclotron frequency of particles with charge $q=Qe$ and mass $m=A m_p$, where $m_p$ is the mass of the proton, $\Omega_{\rm c} = \frac{Q}{A} \Omega_{\rm cp}$. $Q/A=1$ (protons), $2/3$ ($^3$He), 1/2 ($^4$He and heavier elements, if they are fully ionised). In the corona, H and He are expected to be fully ionised, but the highest charge states of Fe observed in SEP events are $Q \simeq 20$. This corresponds to $Q/A \leq 0.4$. As MHD waves cascade towards smaller spatial scales, i.e. increasingly high frequencies, they are hence expected to first resonate with ion species having low cyclotron frequencies, which may then be expected to be more efficiently accelerated than species with higher $Q/A$ ratio. This expected behaviour fits well the observed ordering of enhanced abundances by the $Q/A$-ratio in small impulsive SEP events displayed in Fig.~4.4 of \cite{Rea1999}. For this reason stochastic acceleration is considered as the archetypal ``flare acceleration" process in the SEP literature.

However, the abundance enhancements due to the cascade of MHD waves cannot account for the highly efficient acceleration of $^3$He, which has a high cyclotron frequency. Interpretations of $^3$He enrichments include the resonance with waves produced by electron beams \cite{Mil:al-97,Asc-02} or the presence of multiple resonances of $^3$He with waves when the $^4$He population is accounted for in the dispersion relations of the waves \cite{Pet-12}.

\subsection{Shock acceleration}
Shocks are ubiquitous in the corona, including CMEs and jets. Shock acceleration mechanisms are reviewed in \cite{Jon:Ell-91,KrV-10,Lee:al-12,Des2016}. Below we briefly discuss shock drift acceleration, which is mainly expected at quasi-perpendicular shocks (i.e. in regions of the shock where the normal on the shock front makes a large angle with the upstream magnetic field), and diffusive shock acceleration.

\subsubsection{Shock-drift acceleration}

In the shock-drift acceleration process particles gain energy from the convective electric field $\vec{E} = -\vec{V} \times \vec{B}$, where $\vec{V}$ is the inflow speed of the plasma, $\vec{B}$ the upstream magnetic field in the rest frame of an oblique shock. In the upstream region, electrons and ions $\vec{E} \times \vec{B}$-drift towards the shock front. Because the magnetic field is compressed by the oblique shock, they undergo a gradient drift along the shock front. This drift is directed along the electric field for positively charged particles, and opposite to the electric field for negatively charged particles. Hence electrons, protons and ions gain energy. Depending on the energy and pitch angle before the first encounter with the shock, the energy gained from the drift along the convection electric field may be such that the particle is again injected into the upstream medium and may escape. Besides a gradient drift, particles also undergo a curvature drift while their guiding centre travels along the magnetic field, which is curved in the shock transition. In a planar fast shock, the angle between a field line and the shock normal is larger downstream than upstream, and the curvature drift is opposite to the gradient drift. The curvature drift hence leads to energy losses. But the drift speed decreases as the shock becomes more oblique, so that in quasi-perpendicular shocks the gradient drift, and hence the energy gain, dominates. The acceleration process is widely advocated for electrons in type~II radio bursts \cite{Hol:Pes-83,Man:Kla-05}. But it also applies to SEP-acceleration, especially when particles encounter the shock repeatedly, for instance when they are trapped in the upstream region of a shock that propagates into increasing ambient magnetic field strengths or in a curved ambient magnetic field \cite{San:Vai-06}.

\subsubsection{Diffusive shock acceleration}

When ions are reflected at shock waves and stream into the upstream region, they have a beam-like distribution and are therefore likely to generate waves, parallel-propagating Alfv\'en waves as well as obliquely propagating fast magnetosonic waves. When these waves grow to sufficient amplitudes, they can scatter subsequent ions back to the shock. Since the shock propagates faster than these waves, ions find themselves confined between approaching scattering centers downstream and upstream, and gain energy by bouncing back and forth through the shock font, until they eventually escape. The process requires that particles be able to escape into the upstream medium after the initial reflection to interact with the waves. This means that they must stream away from the shock front at a minimum speed $V / \cos \theta_{\rm Bn}$, where $V$ is the speed of the shock, and $\theta_{\rm Bn}$ the angle between the shock normal and the upstream magnetic field vector. The diffusive shock acceleration process is therefore expected to work best at quasi-parallel shocks. Numerical simulations indicate that it can accelerate protons to relativistic energies \cite{Afa:al-15}. When the shock become more oblique, an increasing fraction of the reflected particles will be unable to escape, and will be convected into the downstream region and thermalised. This is the reason why the injection energy into the diffusive acceleration process is expected to increase with increasing shock angle.

On the other hand, quasi-perpendicular shocks accelerate particles faster and to higher energies. There is hence a theoretical distinction between (1) fast acceleration to high energies at quasi-perpendicular shocks, out of a suprathermal seed population, and (2) slower acceleration to lower energies at quasi-parallel shocks, which can accelerate ions out of the thermal background. In simple scenarios predominantly perpendicular shocks are expected to exist on the flanks of a CME, while shocks near the nose are expected to be quasi-parallel when the surrounding magnetic field is more or less radial, i.e. almost everywhere in the high corona and the interplanetary medium (see Fig.~4 of \cite{Rll:al-16}). However, realistic shocks are expected to have rippled surfaces so that local geometries may be quasi-parallel or quasi-perpendicular irrespective of the location on the shock front.   

\subsubsection{Seed particle population}

The idea that a CME-driven shock would accelerate particles out of the ambient coronal and interplanetary plasma was a prominent justification for the bimodal impulsive-gradual SEP scenario. It seems a natural explanation why the elemental abundances and Fe charge states in gradual SEP events were about coronal at energies up to about 1~MeV/nuc ({\it e.g.}, \cite{Rea1999}), which was the limit of the energy range accessible to in situ measurements of abundances and charge states at that time. This line of reasoning was challenged subsequently by the observation that both charge states and abundances of gradual SEP events depend on energy, and come closer to those observed in impulsive SEP events at energies above 10~MeV (see Sect.~\ref{sec.keyobs}). One way of addressing this problem within a model of shock acceleration is to invoke seed populations that already contain the overabundant elements and the high charge states. In a study of impulsive SEP events from several active regions, \cite{WanYM:al-06} noticed numerous releases of $^3$He-rich particle populations related to coronal jets, and observed enhanced levels of $^3$He throughout the two to three days during which the spacecraft was magnetically connected with the parent active regions. This finding corroborates the idea \cite{Mas:al-99,Mew:al-12} that coronal acceleration episodes during flares or minor activity populate the high corona with enhanced levels of ``flare suprathermals" out of which CME-shocks accelerate particles on occasion to the high energies observed in gradual SEP events. The observed abundances and charge states  would then rather reflect the seed population than the acceleration process.

The argument  that quasi-perpendicular shocks need a seed population with higher energy than quasi-parallel shocks was used to explain energy-dependent differences in the composition between large SEP events \cite{Tyl:Lee-06}. In these authors' view quasi-perpendicular shocks accelerate suprathermal seed populations injected by previous flares, while quasi-parallel shocks would rather accelerate particles out of the thermal background. This clear distinction depends much on the pre-existing turbulence in the acceleration region. \cite{Gia-05} argues that a sufficiently high level of turbulence allows particle acceleration out of the thermal background both at quasi-parallel and quasi-perpendicular shocks. The injection problem as a function of shock angle is discussed, {\it e.g.}, in \cite{Cap:al-15} and \cite{Des2016}.

%%%%%%%%%%%%%%%%%%%%%%%%%%%%%%%%%  Timing relationships between interacting particles and SEPs %%%%%%%%%%%%%%%%%%%%%%%%%%%%%%%%%

\begin{figure}
 \centering{
  \includegraphics[height=6cm,clip=]{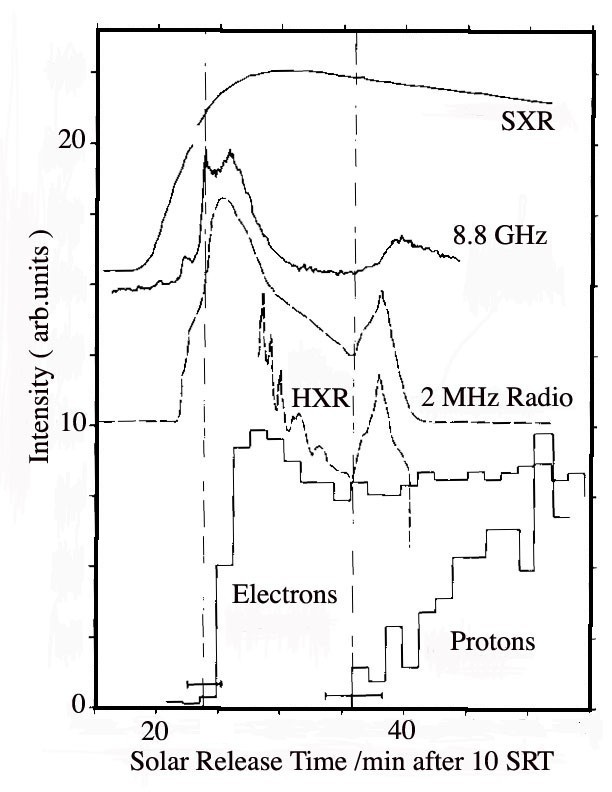}
 \includegraphics[height=4.2cm,clip=]{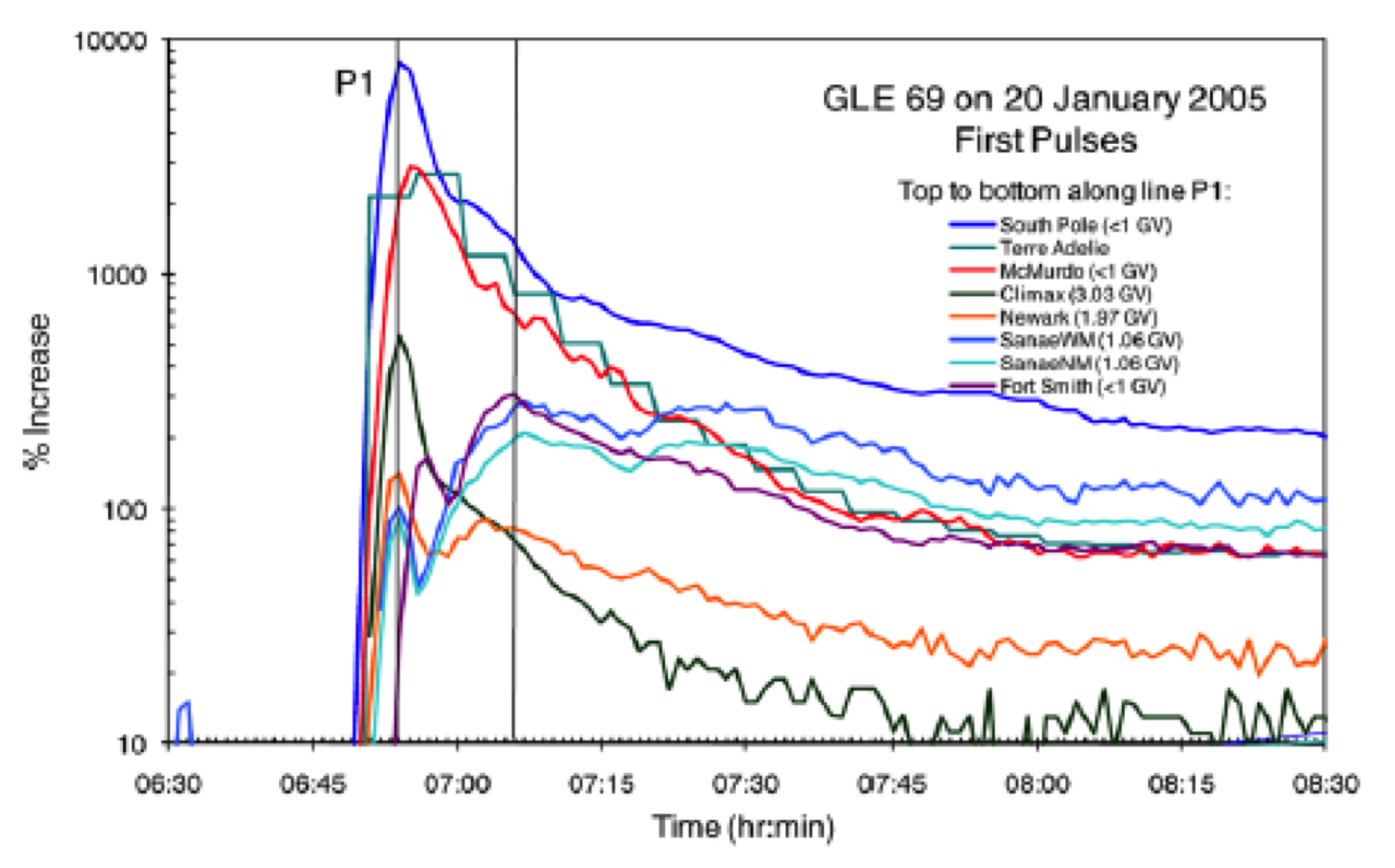}
 }

\vspace{-2ex}
\noindent { (a) \hspace{0.4\textwidth} (b) \hspace{0.48\textwidth} \mbox{} }
\caption{Time structure of SEP events: (a) {\it Helios} SEP measurements (electrons 300-800 keV, protons 27-37 MeV) and associated radio and X-ray emission (HXR: hard X-rays, SXR: soft X-rays) in the corona on 1980 June 08. The abscissa (``solar release time") gives the time when the particles (photons, electrons, protons) were released at the Sun (adapted from \cite{Kal:Wib-91};  \copyright AAS, reproduced with permission). (b) Neutron monitor profiles measured during he GLE on 2005 January 20 \cite{Mor:McC-12}. \copyright Springer. Reproduced with permission.}
\label{Fig_KW} 
\end{figure}

\section{Timing relationships between interacting particles and SEPs}
\label{sec_timing}

\subsection{A solar energetic particle event observed by HELIOS near 0.4 AU}
\label{sec_timing_HELIOS}

The most direct indication of a physical relationship between SEPs in space and non thermal particles interacting with the solar atmosphere is the identification of common time structures. But SEP time profiles observed near 1~AU are usually smeared out by transport in the turbulent interplanetary magnetic field ({\it e.g.}, \cite{Dro-00} - see Sect.~\ref{sec.transp}). \textit{Helios} observed a few SEP events while within 0.4~AU from the Sun. One of the rare events with a well-defined time structure is shown in Figure~\ref{Fig_KW}a (from \cite{Kal:Wib-91}). It was observed when the spacecraft was 0.38~AU from the Sun.  The abscissa gives the time when the particles were released at the Sun, computed by subtracting $\sim 500$~s from the photon arrival time at Earth and the travel time time of the particles along the nominal Parker spiral to \textit{Helios}. The two curves in the bottom panel show that deka-MeV protons started to be released about 10 min after near relativistic electrons. The apparent new rise in the electron profile at that time may be due to cross-talking protons. The important feature is that the releases of the two particle populations are accompanied by distinct episodes of particle acceleration in the corona, as shown by the double-peaked profile of the radio emission at 8.8~GHz, which is gyro-synchrotron emission of near-relativistic electrons (100~keV to a few MeV) in flaring loops in the low corona. The hard X-ray time profile shows a similar structure, but the SMM spacecraft was in the Earth's shadow when the burst began. The radio profile at 2~MHz (ISEE 3) has also two bursts (type III bursts). They are produced by electron beams streaming along open magnetic field lines to the interplanetary space. The simultaneous occurrence of microwave bursts from electrons accelerated in a flare and type III burst hence demonstrates that electrons accelerated during the two microwave bursts can escape from the flaring active region. The succession of an initial electron-rich SEP phase and a second proton-rich  phase is in this case related to two distinct episodes of electron acceleration in the corona. The first, electron-rich phase is connected to a burst during the rise of the soft X-ray burst, shown in the top curve of Figure~\ref{Fig_KW}, hence to the impulsive flare phase. The proton-rich acceleration episode occurs after the soft X-ray peak, in the so-called gradual flare phase. The pronounced difference in the e/p-ratio suggests that the two episodes of particle acceleration did not occur in the same region or involved different mechanisms, or both. The association of each of the two SEP releases with a microwave and a hard X-ray burst argues for closely related particle acceleration processes occurring in the flaring active region and its surroundings. 

\subsection{Time structure of SEPs and transport modelling}
\label{sec_timing_transport}

When observed near 1~AU, particle events with little pitch angle scattering in the turbulent interplanetary magnetic field are rare. But occasionally electrons or relativistic protons may display long mean free paths (see Fig. 4 in \cite{Dro-00}). Agueda and collaborators \cite{Agu:al-08} developed a model of interplanetary electron transport where the time profiles of the intensity and anisotropy of electrons at energies of a few tens to a few hundreds of keV observed in space are reproduced by a series of elementary releases in the corona, at a solar radius above the photosphere. The 1D interplanetary transport model considers focussing along the magnetic field by the overall decrease of the magnetic field strength with increasing heliocentric distance, and pitch angle scattering by the interplanetary magnetic field turbulence. These authors showed that short, impulsive SEP events are well represented by short releases of near-relativistic electrons at the Sun. The inferred releases are accompanied by type~III bursts at decametre-to-kilometre wavelengths, which reveal electron beams travelling from the high corona to 1~AU. In electron events of long duration the situation is more complex, with electron release near the Sun that can last several hours  \cite{Agu:al-14}. These long-duration releases are not accompanied by type~III bursts, but by long-duration broadband bursts at dm-m wavelengths (type~IV), which are presumably related to reconnection in current sheets behind the rising CME, and by type~II bursts related to shock waves in the corona (metre wavelengths) and interplanetary space (kilometre wavelengths). 

Evidence of successive distinguishable SEP releases within a given event has been discussed in recent systematic analyses of ground-level events (GLEs), which are produced by relativistic protons with typical energies in the few GeV range. These primary particles generate atmospheric nuclear cascades that can be detected in ground-based particle detectors, neutron monitors, neutron telescopes or muon telescopes. GLEs often have a double-peaked structure, with an initial fast rise and an anisotropic particle population - called the `prompt component' - followed by a more gradual and less anisotropic `delayed component' (see chap. 7.3 of \cite{Mir-01}). \cite{McC:al-12} demonstrate that the sequence of an anisotropic impulsive peak, and a less anisotropic gradual peak occurring 7-15~min later is a common occurrence when the parent active region is magnetically connected to the Earth, while the absence of the impulsive peak is typical of poorly connected activity near to or east of the central solar meridian or well beyond the western limb.

A prominent case illustrating this double-peaked structure is the GLE of 2005 Jan 20. It displayed a well-defined, rapidly rising time profile at the beginning and a distinct second peak a few minutes later. The difference is clearly seen in the neutron monitor time profiles plotted in Figure~\ref{Fig_KW}b. In a detailed timing comparison, \cite{Kle:al-14} identified microwave and type III counterparts to the two proton peaks similar to the case of 1980 June 08 in Figure~\ref{Fig_KW}a. The second GLE peak was in addition found to occur during a metre wave type IV burst, which is generally ascribed to electron acceleration in the post-CME current sheet. Using the time profile of the microwave burst as the injection profile of relativistic protons at the Sun and modelling the interplanetary transport of the protons in the focussed transport model developed by \cite{Agu:al-08}, \cite{Kle:al-15} showed that a consistent timing could be found with the neutron monitor profiles. This gives support to the idea that different acceleration processes in the low corona successively release the relativistic protons that are observed at Earth. An alternative possibility, advocated by \cite{Mor:McC-12} for the GLE and by \cite{Kal:Wib-91} for the \textit{Helios} observations, is the successive acceleration of the first SEP population in the impulsive flare phase and of the second SEP population at the CME shock. In this scenario it is less clear why there is a timing relationship between SEP acceleration at the CME shock, which at that time is at a heliocentric distance of a few solar radii ({\it e.g.}, \cite{Kle:al-14}), and radiative signatures of renewed electron acceleration in the corona behind the CME. More complex shock acceleration scenarios in the corona were devised that could account for such a situation \cite{Pom:al-08}. But the direct contribution from acceleration processes during magnetic restructuring of the low corona that are expected in the impulsive flare phase, when the CME lifts off, and in the the post-CME corona, is a plausible alternative in the light of these observations. The escape of particles accelerated in a current sheet behind the CME is not fully understood. Reconnection of the closed CME structures, supposedly containing the accelerated particles, with neighbouring open field lines could be essential here, as modelled by \cite{Msn:al-13}.

\subsection{Relativistic nucleons in the solar atmosphere}
\label{sec_timing_FERMI}

The consistent timing between post-impulsive radio or hard X-ray emissions produced by energetic electrons and SEPs suggests that protons are accelerated in the post-CME corona, sometimes up to relativistic energies. Events where radiative diagnostics of energetic ions were reported \cite{Kan:al-93,Koc:al-94,Aki:al-96,Hud:Rya-95} have long been considered as exceptional, because the detectors of nuclear gamma-ray emission were not sensitive enough or were contaminated by protons after the impulsive flare phase. FERMI/LAT changed this situation in discovering that high-energy gamma-ray emission, attributed to the decay of pions produced by protons with energy above 300~MeV, is by no means an extreme phenomenon. Pion-decay gamma-rays were detected in a number of GOES M class flares. On occasions this emission is observed during several hours \cite{Ack:al-14}. The persistence of the gamma-ray source in the vicinity of an active region shown by some imaging observations \cite{Aje:al-14} during long-duration emission is  consistent with acceleration in the low corona. Stochastic acceleration was advocated by \cite{Hud:Rya-95}. However, pion-decay gamma-rays were also observed from flares behind the limb \cite{PsR:al-15a}. This shows that relativistic protons can radiate far from the parent active region. This observation was ascribed to shock acceleration high in the corona. One then needs to understand how the shock maintains a magnetic connection with a localised region in the low solar atmosphere while travelling over many solar radii out into the heliosphere, and how the downward streaming protons overcome magnetic mirroring. The origin of the high energy gamma-ray events, especially those lasting several hours, and their relationship with SEPs is not yet understood.

%%%%%%%%%%%%%%%%%%%%%%%%%%%%%%%%%%%%%%%  Statistical relationships with solar activity} %%%%%%%%%%%%%%%%%%%%%%%%%%%%%%%%%%%%%%

%\section{Observed evidence on SEP sources}
\section{Association between SEP events and solar activity}
\label{sec_stat}

Another attempt to identify the solar origin of SEP events uses statistical relationships. SEP events were only ascribed to flares at a time when CMEs had not been discovered as an independent phenomenon. Since the 1980s CMEs took an increasing importance in the discussion about the origin of SEP events. The ability of CME-driven shocks to accelerate particles all along their way from the Sun to the Earth and beyond is a natural explanation of enhanced SEP intensities that persist over more than a day at energies up to a few tens of MeV (see \cite{Kah-92,Rea1999}). In an attempt to identify candidate solar events that might produce space-weather relevant SEP events, we discuss in the following whether there are preferential associations or statistical correlations between SEP events and either CMEs or flares.  

\subsection{Confined and eruptive solar activity}

Fast CMEs tend to be accompanied by intense flares and vice versa. However, about 10\% of the X class solar flares of cycle 23 were not accompanied by CMEs \cite{Wng:Zhn-07}, they were so-called confined flares. It turned out that these X-ray bursts were not accompanied by SEP events either, even those which were located in the western solar hemisphere, and which therefore were {\it a priori} magnetically connected with the Earth \cite{Kle:al-10,Kle:al-11}. Besides CMEs, the confined flares also lacked radio counterparts at long wavelengths (metre waves and longer), which indicates that flare-accelerated electrons did not get access to extended magnetic structures in the corona and open field lines towards the Heliosphere. Since the particles accelerated in and around the flare remained confined, and since no CME or CME shock was associated with these flares, no SEP event could be expected. This is confirmed by the impressive series of X-class flares without CMEs that was observed from one active region during its disk passage in October 2014 \cite{Tha:al-15}. It produced six X-class flares (X1.0-X3.1), among them four while in the western solar hemisphere (October 22-27), with no CME. The GOES particle intensity measurements did not exceed background during this time (\url{http://www.solarmonitor.org}).  As shown in \cite{Tha:al-15}, the flares occurred in an environment with strong surrounding and overlying magnetic fields, like in the events analyzed in \cite{Wng:Zhn-07}. Electron acceleration accompanied the flares, as shown by the hard X-ray emission. But none of the X-class flares was accompanied by conspicuous type~III bursts at decameter and longer waves (Wind/WAVES observations), and none of the three bursts that occurred during observing hours of the Nan\c{c}ay radio observatory was accompanied by a radio burst between 1~GHz and 30~MHz (see \url{http://radio-monitoring.obspm.fr}). 

\subsection{SEP events without flares?}

So while confined flares as defined above do not produce SEP events, one can ask the opposite question: do CMEs that have no other particle acceleration signature in the corona, such as hard X-rays or radio emission at cm-to-dm wavelengths, produce SEP events? Kahler and coworkers \cite{Kah:al-86} studied a filament eruption and CME without an impulsive flare or radio emission from the low corona, but which was associated with  a type~II burst at hectometric wavelengths, i.e. a shock wave at heliocentric distances beyond a few solar radii. The CME was accompanied by an SEP event with a steep spectrum, seen up to about 80~MeV, and also by near-relativistic electrons. This is widely considered as a prime example that a CME shock is a sufficient condition to produce SEPs, although it was also pointed out  \cite{Can:al-02} that besides the type II burst a type III burst, signalling electron acceleration in the lower corona, was also observed in association with a weak SXR burst. Whatever the interpretation of the filament-associated SEP events, there are at best very few SEP events associated with a CME and no alternative signature of particle acceleration in the corona (see also \cite{Mar:al-06}).

\subsection{Statistical relationships between SEP peak intensities and parameters of solar activity}

Besides considering the common occurrence of SEP events, flares and CMEs, one can search for a preferential correlation between parameters describing the importance of the SEPs on the one hand and of the eruptive solar activity on the other hand. Many years of work led always to the same significant, but noisy type of correlation between the SEP peak intensity, at energies of a few MeV to a few tens of MeV, and the peak flux of the SXR burst on the one hand, the CME speed on the other ({\it e.g.},  \cite{Kah-01,Gop:al-03,Can:al-10,Mit:al-13,Ric:al-14}). Comparable correlations were found with CME speed and with SXR peak flux. A correlation between the flux of protons above 10 MeV in space and the gamma-ray fluence in the 4-7~MeV range, where nuclear lines are superposed on the bremsstrahlung continuum, was found in \cite{Chr-90}.

Kahler \cite{Kah-01} pointed out that the correlation between SEP intensity and CME speed could be blurred by the contribution of other parameters. He found that higher peak SEP intensities were observed when the pre-event background was higher. He argued that this was evidence that high SEP intensities result when there is a suprathermal seed population for shock acceleration in the high corona and the interplanetary space. CME speed by itself is also not as useful a parameter to characterise the strength of a shock as is the Mach number, which cannot be measured directly in the corona. In a systematic study of fast CMEs, Gopalswamy and coworkers \cite{Gop:al-08} showed that those which lacked type II emission at decametric and longer waves had no SEP event. The authors concluded that this pointed to a varying Alfv\'en speed in the corona, and that even CMEs with speeds as high as 1000~km~s$^{-1}$ did not necessarily drive a shock. The existence of a type~II burst, however, proved the presence of the shock, and strengthened the idea that the shock revealed by the radio emission also accelerated the SEPs.  Clearly, the conventional correlation analysis oversimplifies the problem.

Another reason to explain the failure of classical correlation studies in establishing preferential associations of SEP importance with flare importance or CME speed is the fact that the solar activity parameters are themselves correlated \cite{Kah-82b,Tro:al-15}. More sophisticated statistical methods then suggest that SEP peak intensities are indeed correlated significantly both with CME speed and with flare parameters, especially SXR fluence \cite{Tro:al-15}. This argues for a mixed scenario, where processes related with both flares and CMEs contribute to SEP acceleration.

This conclusion is confirmed by a detailed study of the dependence on SEP energy. Most statistical studies were limited to a single energy range. When considering SEP peak intensities in different energy ranges between 1 MeV and 100~MeV, Dierckxens and coworkers \cite{Drc:al-15}  discovered a systematic trend: at energies well below 20 MeV, the correlation coefficient with the SEP peak intensity is higher for CME speed than for SXR peak flux, but at energies above 20~MeV the inverse is true. This is new evidence for a mixed acceleration scenario, where different acceleration processes dominate at different SEP energies. This is consistent with the result discussed above that SEP events associated with the eruption of quiescent filaments have steep energy spectra \cite{Kah:al-86,Gop:al-15}.

\subsection{Comparisons of the numbers of particles interacting in the solar atmosphere and particles in SEP events at 1~AU}

Quantitative comparisons between the spectra and numbers of interacting particles in the solar atmosphere and SEPs are important further clues, but depend strongly on the models employed. Collisional bremsstrahlung at hard X-rays and nuclear gamma-ray emission are in principle understood, and can be modelled. But simple models for the energy spectrum, the angular distribution, and the nuclear abundances of the non-thermal and target particle populations have to be assumed. In SEP events, single-point measurements often have to be used to estimate SEP intensity over an extended range in heliocentric longitudes and latitudes, and the total number is inferred using an assumption of the spatial extent. While multi-spacecraft observations bring supplementary constraints \cite{Lar:al-13,Dre2014}, one still has to suppose a simple smooth variation of the particle intensity with heliolongitude. This may be unrealistic, if particles are released into discrete channels of the interplanetary space, implying an irregular variation of the intensity with spacecraft position, as shown for protons by flux dropouts in impulsive SEP events \cite{Maz:al-00}, and for electrons of low energy emitting type III radio bursts \cite{But-98} as well as for near-relativistic electrons \cite{Kla:al-16}. So any quantitative comparison between interacting protons and SEPs at 1~AU cannot be done to better accuracy than an order of magnitude.

Such comparisons were conducted for interacting protons and protons in SEPs at energies of a few tens of MeV. In a study of several events of the SMM era, Ramaty and coworkers \cite{Ram:al-93} found an event-to-event variation of the ratio of protons at energies above 30 MeV inferred from gamma-ray modelling and from in situ measurements in a broad range between about 0.01 and 100 (see also \cite{Ems:al-05}). The number of protons above 30~MeV in the large SEP event on 1990 May 24 was found to be smaller than the number of protons interacting in the low solar atmosphere \cite{Deb:al-97}. For electrons in the energy range between a few tens of keV and a few MeV the situation is different: the number of electrons detected in space is always found to be orders of magnitude smaller than the number required for hard X-ray emission by collisional bremsstrahlung, although a correlation seems to exist between the energy spectra of the two populations \cite{Ram:al-93,Dro-96,Kru:al-07}. 

%%%%%%%%%%%%%%%%%%%%%%%%%%%%%%%%%%%%%%%   SEP transport %%%%%%%%%%%%%%%%%%%%%%%%%%%%%%%%%%%%%%% %%%%%%%%%%%%

\section{Particle transport}   \label{sec.transp}

Once acceleration has taken place, propagation from the source region distributes particles within the heliosphere, and in some cases they can reach near-Earth space, where they may produce Space Weather effects. The magnetic fields of the corona and interplanetary space are key to transport. The coronal fields have a complex geometry including closed and open magnetic field configurations. The average interplanetary magnetic field can be described as a Parker spiral, superimposed on which are a variety of transient structures as well as turbulence. The latter produces scattering of energetic particles and meandering of the magnetic field lines.  

\subsection{Coronal magnetic field}

The longitude distribution of parent flares of SEP events peaks at Western heliolongitudes (Figure~\ref{Fig_ldist}), as one would expect considering that the footpoint of the Parker spiral for an Earth observer is located in this region. However, the distributions are broad, and multi-spacecraft observations discussed in Sect.~\ref{sec.keyobs} show that in individual cases the longitudinal injection cones may exceed 180$^\circ$. 

The width of the heliolongitude distribution tends to be larger than the range over which the heliolongitude of the Earth-connected Parker spiral on the solar wind source surface varies due to varying solar wind speed. One important contribution to broad particle injection cones is the geometry of the coronal magnetic field above active regions. While active regions are largely composed of closed magnetic fields, there are also open field lines. A well-known illustration of this fact are type III radio bursts, which are produced by electron beams travelling outward throughout the corona and the interplanetary space, and are a very frequent form of radio emission related with solar flares ({\it e.g.}, \cite{StH:al-13}). Other evidence comes from the extrapolation of photospheric magnetic field measurements to the source surface using models such as the Potential-Field-Source-Surface (PFSS) model of \cite{Scr:DeR-03}. The open field lines form very narrow bundles at altitudes below, say, 0.5~R$_\odot$ above the photosphere, due to the pressure of neighboring closed magnetic field structures. But the closed magnetic flux diminishes with increasing height, and the open field lines are bound to spread, connecting a tiny part of an active region with a longitude range as large as 90$^\circ$ at the source surface ({\it e.g.}, \cite{Lie:al-04}). This spreading was conjectured in the interpretation of early SEP observations \cite{Fan:al-68,Rei:Wib-74}. Using radio observations of electron beams in the corona, \cite{Kle:al-08} showed that it explains the observation of simple electron events at Earth related to parent activity as far as 50$^\circ$ away from the nominal Parker spiral longitude. It is, however, not sufficient to explain very broad injection cones observed in some simple impulsive SEP events observed by STEREO \cite{Wie2013}. But the spread of open magnetic field lines cannot be ignored in the discussion of particle transport in the corona.

\subsection{Classical 1D models of interplanetary transport}

The classical description of particle transport in the interplanetary medium relies on a 1D focussed transport equation, describing the effects of focussing along the magnetic field and scattering due to magnetic field fluctuations \cite{Roe-69}. The only spatial variable retained in the model is the distance measured along a Parker spiral magnetic field line, so that particles are assumed to remain bound to the field line on which they were injected, for their entire propagation. A revised formalism that includes a description of adiabatic deceleration was later developed \cite{Ruf1995} and this forms the basis of several current SEP models. 

Within a 1D formalism, the value of the scattering mean free path $\lambda$ is a key parameter in determining the overall shape of SEP intensity profiles at 1 AU. Many studies used the focussed transport equation to fit measured intensity profiles (and in some cases anisotropies) and derive a value of $\lambda$ ({\it e.g.},~\cite{Kal1993}). These types of studies generally found regimes of strong scattering for protons, with typical parallel mean free paths of $\sim$0.1 AU, with weak dependence on rigidity \cite{Bie1994}. For electrons, the radial mean free path is typically between 0.1 and 0.5 AU, showing a tendency to decrease with increasing particle rigidity \cite{Agu:al-14}. 

The 1D assumption is embedded within the two-class paradigm, in which  the wide extent in longitude of many large events is explained as efficient acceleration over a wide portion of a CME-driven interplanetary shock \cite{Rea1999}. The role of scattering in shaping intensity profiles of SEPs accelerated by interplanetary shocks is less clear than for the case of a compact injection region at the Sun. For the former case, in fact, acceleration can continue as the shock has left the corona, resulting in a time extended injection. Models that aim to account for an extended injection from a shock as well as propagation, need to consider that a given observer will be magnetically connected to different parts of the shock over time \cite{Lar1998,Bai2016}.

\subsection{3D models of interplanetary transport}

Interaction with turbulence in interplanetary space may produce particle diffusion across the magnetic field, in addition to the effects on propagation along the field, already included in classic 1D models.

For this reason, extensions of the focussed transport equation to include scattering across the magnetic field have been developed \cite{Zha2009,Dro:al-10}. Here particles may slowly diffuse away from the magnetic field line on which they were injected, requiring a 3D description. Parallel and perpendicular mean free paths are an input to the models.

The need for 3D modelling was also emphasized by studies based on the test particle approach, where the full equations of motion are integrated numerically for an energetic particle population. When this methodology was applied to SEPs for the case of a Parker spiral configuration, it was found that guiding centre drifts associated with the gradient and curvature of the magnetic field produce transport in both longitude and latitude, in the absence of perpendicular scattering \cite{Mar2013,Dal2013}. One important consequence of drift is the associated deceleration \cite{Dal2015}.

Drifts are particularly strong for SEP heavy ions due to their large $m/q$, which results from being partially ionised. When heavy ions are injected from a compact area at the Sun, they are found to propagate efficiently to locations that are not magnetically well connected to the injection region. This drift-dominated propagation produces a decrease over time of the Fe/O ratio \cite{Dal2016a} similar to that observed in SEP events \cite{Mas2006}, for not magnetically well connected observers. The reason for the observed decrease in Fe/O is that Fe drifts more than O due to its larger mass-to-charge ratio, so that events tend to be Fe rich early on.

In addition, a 3D model including drift naturally produces an energy dependence of the 1 AU charge states for an observer without direct connection to the injection region \cite{Dal2016b}, qualitatively similar to SEP observations \cite{Maz1999}. In fact, due to the dependence of drift velocity on the product of $m$/$q$ by energy per nucleon, for particles at low energy a low charge state is the only way to ensure significant drift across the field. At higher particle energies, higher charge states are also able to drift efficiently, resulting in a larger measured event charge state. The timescales over which drift effects become important depend on the specific particle species, energy and charge-to-mass ratio. For heavy ions they are important over timescales of the order of an hour.

In addition to scattering of energetic particles, turbulence in the interplanetary magnetic field produces meandering of the magnetic field lines, so that the magnetic connection between a given location in space and the source of an SEP event may differ considerably from what is predicted by a simple Parker spiral model. \cite{Lai2016} incorporated field line meandering into a model of SEP 
propagation in the heliosphere and showed that it can explain the fast onset of SEP events over wide longitudinal ranges, as can be observed experimentally. They also showed that the early SEP propagation across the magnetic field cannot be described by a diffusive approach.

%---------------------------
%%%%%%%%%%%%%%%%%%%%%%%%%%%%%%%%%%%%%%%   SEP forecasting %%%%%%%%%%%%%%%%%%%%%%%%%%%%%%%%%%%%%%%%%%%%%%%%%%
\section{Approaches to SEP forecasting}  \label{sec.forecast}

Forecasting SEP events has attracted a large amount of effort in recent years. Producing forecasts with long lead times (of the order of days) relies on the analysis of solar active regions and their magnetic configuration. Short lead time forecasts, on the other hand, are issued when a solar eruptive event, typically including a solar flare and/or a CME, is detected. Here we focus on the latter scenario, where the aim is to predict whether or not the observed eruption will produce SEPs at Earth, and in the case of positive forecast to estimate the impact the SEP event is likely to have. In the following some recent models are briefly described. An earlier overview is given in \cite{Vai:al-09}. 

\subsection{Empirical models}

Empirical models are based on known statistical relationships between various types and magnitudes of solar events and SEPs, as reviewed in Section \ref{sec_stat}. 
The PROTONS system, the SEP prediction tool in use at the NOAA Space Weather Prediction Center (SWPC), makes uses of an algorithm that takes into account the GOES X-ray peak flux of the flare, as well as information on the flare location \cite{Bal2008}. Another empirical tool is the Proton Prediction System (PPS), also based on flare parameters \cite{Kah2007}.
The SEP forecasting tool developed within the {\it COronal Mass Ejections and Solar Energetic Particles: forecasting the space weather impact} (COMESEP) alert system makes use of empirical statistical relationships to forecast the occurrence of an event \cite{Drc:al-15}, while it uses a physics based model to forecast event parameters \cite{Mar2015}.
The SEP prediction model of the University of Malaga (UMASEP model) tracks simultaneously the time profiles of soft X-rays and protons observed by the GOES spacecraft \cite{Nun-11}. It predicts the occurrence, peak time and peak intensity of an SEP event. This model has been incorporated within the SEPsFLAREs forecasting system \cite{Gar2016}.
Methods that forecast proton arrival based on the detection of relativistic electrons or relativistic protons, which travel to 1 AU faster than ions, have also been proposed \cite{Pos2007,Svt:al-14}.

Besides the above-mentioned models a number of empirical relationships have been discovered that may be exploited in future operational models.
\begin{itemize}
\item The examination of hard X-ray bursts observed by the {\it Solar Maximum Mission} (SMM) showed that those associated with SEP events exhibit a distinct spectral hardening ({\it i.e.} flattening)  throughout the event \cite{Kip-95}. This is a distinctive feature, since in typical hard X-ray bursts the photon spectrum starts soft, hardens as the intensity rises, and softens again in the decay phase \cite{Gri2004}. The SMM-finding was confirmed by RHESSI observations \cite{Gra:al-09}. 
\item Chertok and coworkers \cite{Chr:al-09} noticed that SEP events at energies above 100~MeV were accompanied by microwave bursts with spectra that had peaks at or above about 15~GHz, whereas on average microwave peak frequencies are around 10~GHz. A possible interpretation is that SEP events are associated with particularly strong electron acceleration, leading to high densities of radio-emitting electrons in the flaring active region.  
\item When examining the soft X-ray bursts associated with SEP events, it was noticed that apparently cooler events, which means events that were relatively faint in the 0.05-0.4~nm channel of the GOES monitoring instrument as compared to the 0.1-0.8~nm channel, are more strongly associated with SEPs than the others (see \cite{Gar-04}, and references therein). 
\end{itemize}
Both the continuous spectral hardening of hard X-ray bursts and the low temperature of the SXR bursts associated with SEP events may translate the association of SEP events with extended coronal structures that exist during CMEs. But it is not clear why continuous hardening of the hard X-ray spectrum indicates escaping SEPs, especially since the effect is usually ascribed to energy-dependent energy losses or energy-dependent precipitation of a trapped electron population.

\subsection{Physics based models}

Physics based models aim to provide an SEP forecast by modelling the relevant acceleration and transport processes. This is complicated by the fact that only limited information about the acceleration is known within a real time forecasting context, so that a number of assumptions are usually necessary to produce predicted SEP intensity profiles.

The {\it Solar Particle Engineering Code} (SOLPENCO) forecasting framework solves a 1D focussed transport equation coupled with MHD modelling of the shock accelerating the SEPs \cite{Ara2006}. Here the fact that over time an observer is connected to different portions of the shock, having varying acceleration efficiencies is taken into account.

An alternative approach uses the output of ENLIL simulations to obtain information on the location and properties of CME driven interplanetary shocks \cite{Bai2016}. Particles injected from these shocks are propagated in 1D along the magnetic field within a scatter-free assumption \cite{Luh2010}.

The {\it Energetic Particle Radiation Environment Module} (EPREM) within the {\it Earth-Moon-Mars Radiation Environment Module} (EMMREM) framework couples a 1D focussed transport equation with a convection-diffusion equation to describe transport in 3D \cite{Sch2010}. 

The SPARX model, based on the test particle approach, solves for SEP trajectories in 3D to forecast time profiles of  particle intensities at 1 AU \cite{Mar2015}.

In recent years, new numerical models have also attempted to couple realistic simulations of CME propagation in the corona with particle simulations to describe the associated particle acceleration \cite{Man2005,Koz2013}.

\section{Discussion} \label{sec.disc}
%---------------------------

The manifestations of energetic particles of solar origin in the Earth's plasma environment are a complex interplay between acceleration at the Sun and in the interplanetary space, and of transport in the turbulent interplanetary magnetic field. As of today it is impossible to predict the occurrence and properties of SEP events from properties of an active region or of the coronal magnetic field before a solar eruptive event occurs. The earliest observable signature related to an upcoming particle event is electromagnetic emission. The earliest electromagnetic emissions are related to the impulsive flare phase. Forecasting schemes of SEP events that use electromagnetic emissions, mostly soft X-rays because of their continuous availability from the GOES monitoring observations, therefore rely on the hypothesis of at least a statistical, if not physical, connection between the energy release in a flare and the importance of the associated SEP event.

Within the two-class scenario for SEP events, where impulsive and gradual events are distinguished, events with the largest space-weather impact are of the gradual type, i.e. they are associated with coronal and interplanetary shocks driven by fast CMEs. It is widely considered that it is these CME shocks that accelerate the SEPs. This conclusion appears to be in conflict with the use of flare-related emission in SEP forecasting schemes. An important question is therefore whether acceleration in flares, that in most cases happens in conjunction with sufficiently fast CMEs, is also a source of space-weather relevant SEPs. Furthermore, the liftoff of CMEs is also followed by the formation of current sheets and by signatures of magnetic reconnection that persist well beyond the impulsive flare phase. So there is a number of possible processes that may contribute to the acceleration of particles released into the interplanetary space. The term ``flare acceleration" should not be restricted to magnetic reconnection processes and turbulence in the impulsive flare phase, but considered in a more general sense. 

The energy range of SEPs spans many orders of magnitude, from hundreds of keV to possibly tens of GeV. Different space weather aspects refer to different particle energy ranges: tens of MeV for the impact on space missions outside the Earth's magnetosphere and in polar orbits, on human spaceflight outside the magnetosphere, and on radio communications through the polar ionosphere of the Earth; $>$1~GeV for the production of atmospheric particle cascades and their impact on aviation. Many observables of SEP events, including the longitude distribution of the parent activity, the elemental abundances and the Fe charge states, depend on the particle energy. As a result, it is not possible to define a sharp separation between `impulsive' and `gradual' SEP events that holds across the entire energy spectrum. The relative contribution of acceleration at CME shocks on the one hand and in small-scale plasma processes  involving magnetic reconnection on the other is hence likely to depend on the energy of the particles being considered. At which energies the distinction is made is presently an open question and needs more research.The work of \cite{Drc:al-15} suggests a stronger dependence of SEP intensities on CME speed than on soft X-ray flux at energies below 10~MeV, and the opposite at energies above 20 MeV. The transition in the range 10-20~MeV is a statistical average, not a sharp physical limit.

The complexity of particle acceleration models has considerably increased in recent years, in order to cope with the observations. The role of fragmentation of the acceleration region in reconnecting current sheets has been emphasised by physical considerations and numerical modelling. As far as shock acceleration is concerned, the importance of the seed population has been highlighted by observations that show sometimes enhanced levels of $^3$He and other elements in the interplanetary medium. Shock acceleration out of this seed population would then create energetic particle populations with elemental abundances that contain the peculiar enhancements of the seed population. This points to the importance of particularly active regions in the production of major solar events, where numerous flares and CMEs occur before the eruptive event at the origin of the SEPs, and which apparently release the suprathermal seed particle populations over times that are much longer than the electromagnetic emissions of the flares themselves. 

But in addition to the question of particle acceleration, understanding the transport of SEPs is the other key to successful forecasting. While classical models of SEP propagation have made use of the 1D approximation, which assumes particles remain tied to the magnetic field line on which they are injected, the importance of using a 3D description has recently been emphasised. Within the standard two-class scenario, properties such as elemental abundances and charge states of heavy ions are ascribed purely to the acceleration process. Describing particle motion in 3D and accounting for the effects of drifts opens up the possibility that such properties may be partly the result of propagation, for the case of an observer that is not well connected to the source. In addition to guiding centre drifts in the spatially varying interplanetary magnetic field, 3D propagation may also be caused by processes such as scattering by turbulence and field line meandering. An accurate description of the complex magnetic fields of the solar corona is also of great importance to understanding SEP escape and propagation.

A number of institutions and projects are attempting to make SEP forecasts and improve their reliability. At present, a number of factors are limiting the efficacy of SEP forecasts. On the observational side, although real time flare observations are well established, little information is available on CME shocks, including their extent and Mach numbers. As a result it is difficult to obtain observational information on SEP injection in a real time forecast scenario. On the physics modelling side, a balance needs to be struck between correctly including the relevant physics of acceleration and transport and computational efficiency, so that useful information can be obtained over timescales relevant for forecasts. 

A major limitation on our understanding of SEP acceleration and propagation is the exclusive availability of in situ measurements near 1~AU. Effects of particle acceleration and particle transport are then mixed and difficult to disentangle. We expect that the upcoming Solar Orbiter and Solar Probe Plus missions will provide invaluable new measurements of SEPs. By going closer to the Sun, these missions will allow detection of SEP properties closer to the acceleration region, therefore minimising any modifications introduced by transport effects. In addition they will be able to detect shocks and the associated populations of particles and plasma waves much closer to the Sun, where their particle acceleration efficiencies are expected to be much higher. The new insight gained on particle acceleration and transport by these missions will enable a major step foreward in our forecasting abilities.

\begin{acknowledgements}
The authors are grateful to ISSI for organizing the workshop on the scientific foundations of space weather, as well as a number of earlier team meetings that helped us shape our ideas on solar energetic particles. S.D.~acknowledges support from the UK Science and Technology Facilities Council (STFC) (grant ST/M00760X/1) and the Leverhulme Trust (grant RPG-2015-094). Research at Paris Observatory received funding from the French space agency \emph{CNES} and the European Union's Horizon 2020 research and innovation programme under grant agreement No 637324 (HESPERIA project). 
\end{acknowledgements}

% BibTeX users please use one of
%\bibliographystyle{spbasic}      % basic style, author-year citations
%\bibliographystyle{spphys}       % APS-like style for physics
%\bibliographystyle{aps-nameyear}

%\bibliographystyle{spmpsci}      % mathematics and physical sciences
%\bibliography{sep_biblio}}   % name your BibTeX data base

\end{document}